\documentclass[reprint,aps,prb,superscriptaddress,showpacs,epsf]{revtex4-1}
\usepackage{float}
\usepackage{color,graphicx}
\usepackage{graphics}
\graphicspath{ {images/} }
\usepackage{hyperref}
\usepackage{mathrsfs}
\usepackage{dcolumn}
\usepackage{amsmath,amssymb}
\begin{document}
\title{Anomalous magnetic properties of quasi two-dimensional van der Waals ferromagnet Fe$_4$GeTe$_2$}
\author{Suchanda Mondal}
\affiliation{Saha Institute of Nuclear Physics, HBNI, 1/AF Bidhannagar, Kolkata 700 064, India}
\author{Prabhat Mandal}
\email{prabhat.mandal@saha.ac.in}
\affiliation{S. N. Bose National Centre for Basic Sciences, JD Block, Sector III, Salt Lake, Kolkata, 700106, India}
\date{\today}

\begin{abstract}
We have studied the magnetic  properties of quasi two-dimensional itinerant ferromagnet Fe$_4$GeTe$_2$ as functions of temperature and magnetic field ($B$) with field parallel to the $ab$-plane ($B\|ab$) and $c$-axis ($B\|c$) of the crystal. This van der Waals compound undergoes a continuous paramagnetic to ferromagnetic  phase transition below  $T_C$ = 270.0 K and a pronounced spin reorientation transition  at around $T_{SR}$ $\sim$ 115 K where the easy axis of magnetization changes its direction from in-plane to out-of-plane. The temperature evolution of magnetization in the FM state is highly anomalous and extremely sensitive to the direction  and strength of applied magnetic field. Magnetic entropy change ($\Delta S_M$) has been estimated in the vicinity of $T_C$  and $T_{SR}$. $\Delta S_M$($T$) is found to be almost isotropic around $T_C$ while it shows very unusual behavior and is sensitive to the direction of applied field  at low temperature close to $T_{SR}$. For $B\|c$, $\Delta S_M$ is negative and decreases continuously with field and -$\Delta S_M$($T$) shows a broad maximum around $T_{SR}$ similar to that observe at $T_C$. On the other hand, $\Delta S_M$ is  positive in the vicinity of $T_{SR}$ below 1.4 T and -$\Delta S_M$($T$) exhibits a peak at $T_{SR}$ which becomes very sharp at high field for $B\|ab$. These results suggest complex nature of the magnetic ground state of Fe$_4$GeTe$_2$, which is possibly due to the presence of multiple inequivalent Fe sites, their ordering and  complexity of spin configuration.
\end{abstract}

\maketitle
\section{Introduction}
The quasi two-dimensional (2D) van der Waals (vdW) family of compounds offer a fertile playground to investigate the novel quantum phenomena \cite{AKGeim,TSong,DKlein,CGong,Mcg,BHuang}. In these materials, the weak coupling between the layers allows easy mechanical exfoliation to control the dimensionality and stacking configuration which create huge possibilities for exploring materials with tailored physical properties. As an emerging member of this family, 2D ferromagnet with intrinsic long-range magnetic ordering, in which the electron spin is used as an additional degree of freedom to bring rich physics and more functionalities, may pave a way to build-up high-performance and energy-efficient devices \cite{Mcg,BHuang,Burch,Casto,Carte,Tsubo,Deng,Deise,Siber,Chen,KKim}. Unlike conventional ferromagnets, the magnetic interaction in vdW ferromagnets is highly anisotropic in nature: the exchange interaction within the $ab$-plane is strong, while the magnetic coupling between the layers is weak. The spin-orbit interaction plays an important role in sustaining the long-range ferromagnetic  (FM) ordering down to monolayer or bilayer at a finite temperature. This unique behavior of vdW ferromagnets is very important to fabricate devices that show various emergent spin-orbit coupled phenomena and different types of symmetry breaking. However, the Curie temperature ($T_C$) of existing 2D ferromagnetic materials is far below the required operating temperature for applications. Thus, it is important to search new materials with high transition temperature, close to room temperature. In recent times, the vdW ferromagnets with general chemical formula Fe$_n$GeTe$_2$ ($n\geq3$) have received considerable attention as they appear to be promising candidates for realizing stable vdW material with metallic conductivity and large magnetization ($M$), which are the key components for the spintronic applications \cite{KKim,YWang,CTan,AFMay1,AFMay2,JSeo1,Deng,Chen,Zhang,Stahl,Gao,Fei,Ding}. In this class of materials, both $T_C$ and physical properties can be controlled by tuning the Fe concentration  and layer thickness. Bulk single crystal of Fe$_3$GeTe$_2$  shows $T_C$ $\approx$ 220 K and the transition temperature decreases with decreasing layer number and reduces to 130 K for monolayer \cite{Chen,Fei}. However, $T_C$ of atomically thin layer can be tuned to 300 K by ionic liquid gating \cite{Deng}.  As a FM nodal line semimetal, Fe$_3$GeTe$_2$ exhibits large anomalous Hall effect and the FM stripe-like domain structure in few layers of this material converts into skyrmion bubble when the magnetic field is applied perpendicular to $ab$-plane \cite{Ding}. This clearly  signifies the exquisite topological spin configuration.

Even though the off-stoichiometric compound Fe$_{5-x}$GeTe$_2$ shows $T_C$ above room temperature, its $T_C$  is sensitive to Fe-vacancy and thermal history \cite{AFMay1,AFMay2,Zhang,Stahl,Gao}. Whereas the newly discovered Fe$_4$GeTe$_2$ compound has been reported to exhibit well defined FM transition close to room temperature (270 K) \cite{JSeo1,Mondal}. With further cooling below $T_C$, Fe$_4$GeTe$_2$ exhibits an additional transition around $T_{SR}\sim$115 K, the spin reorientation  transition \cite{JSeo1,JSeo2,Mondal}. The other members of Fe$_n$GeTe$_2$ family do not show such transition but a definite direction of magnetic anisotropy at all temperatures \cite{AFMay2,Zhang}. In Fe$_4$GeTe$_2$,  this temperature-dependent anisotropy due to the spin reorientation transition leads to a crossover from perpendicular magnetic anisotropy (PMA) to in-plane magnetic anisotropy \cite{JSeo1}. It has been shown that by reducing the thickness of Fe$_4$GeTe$_2$ down to few layers limit, PMA increases and hard ferromagnetic nature enhances \cite{JSeo1}. These behavior indicate that the nature of magnetism in Fe$_4$GeTe$_2$ is very rich and complex due to the spin reorientation transition. However, there is no detailed magnetization measurement  in the vicinity of $T_{SR}$ and its analysis for understanding the intricate nature of the transition.

In this work, we have done systematic study of the magnetic properties of Fe$_4$GeTe$_2$ compound along different crystallographic axes as  functions of temperature and magnetic field. The temperature dependence of magnetization in the FM state shows several anomalies which are very sensitive to the direction and strength of applied magnetic field. We have also estimated the magnetic entropy change ($\Delta S_M $). As a differential physical parameter, $\Delta S_M $ is a very sensitive tool to determine the nature of magnetic ground state and  phase transition.  Indeed, $T$ and $B$ dependence of $M$ and $\Delta S_M $ for Fe$_4$GeTe$_2$ reveal several interesting and complex behavior across the spin reorientation transition. The observed results are compared and contrasted with that reported for other members of this series as well as several other vdW compounds.

\section{Sample Preparation and Experimental Techniques}
High-quality single-crystalline samples of Fe$_4$GeTe$_2$ have been grown by the chemical vapor transport technique.  The details of sample preparation and characterization are given in our earlier report \cite{Mondal}. The temperature and magnetic field dependence of magnetization measurements have been performed in a superconducting quantum interference device-vibrating sample magnetometer (MPMS 3, Quantum Design) and physical property measurement system (PPMS, Quantum Design) with field up to 7 T. Isothermal magnetization data were collected in a slow field sweep mode over a wide range of temperature from 320 K down to 2 K, covering both FM-paramagnetic (PM) and spin reorientation transitions. The measurements were performed after stabilizing the sample temperature for 10 min. $M$($T$) data at several applied fields were recorded in a slow temperature sweep mode.

\section{Results and Discussions}
\begin{figure}
\includegraphics[width=0.5\textwidth]{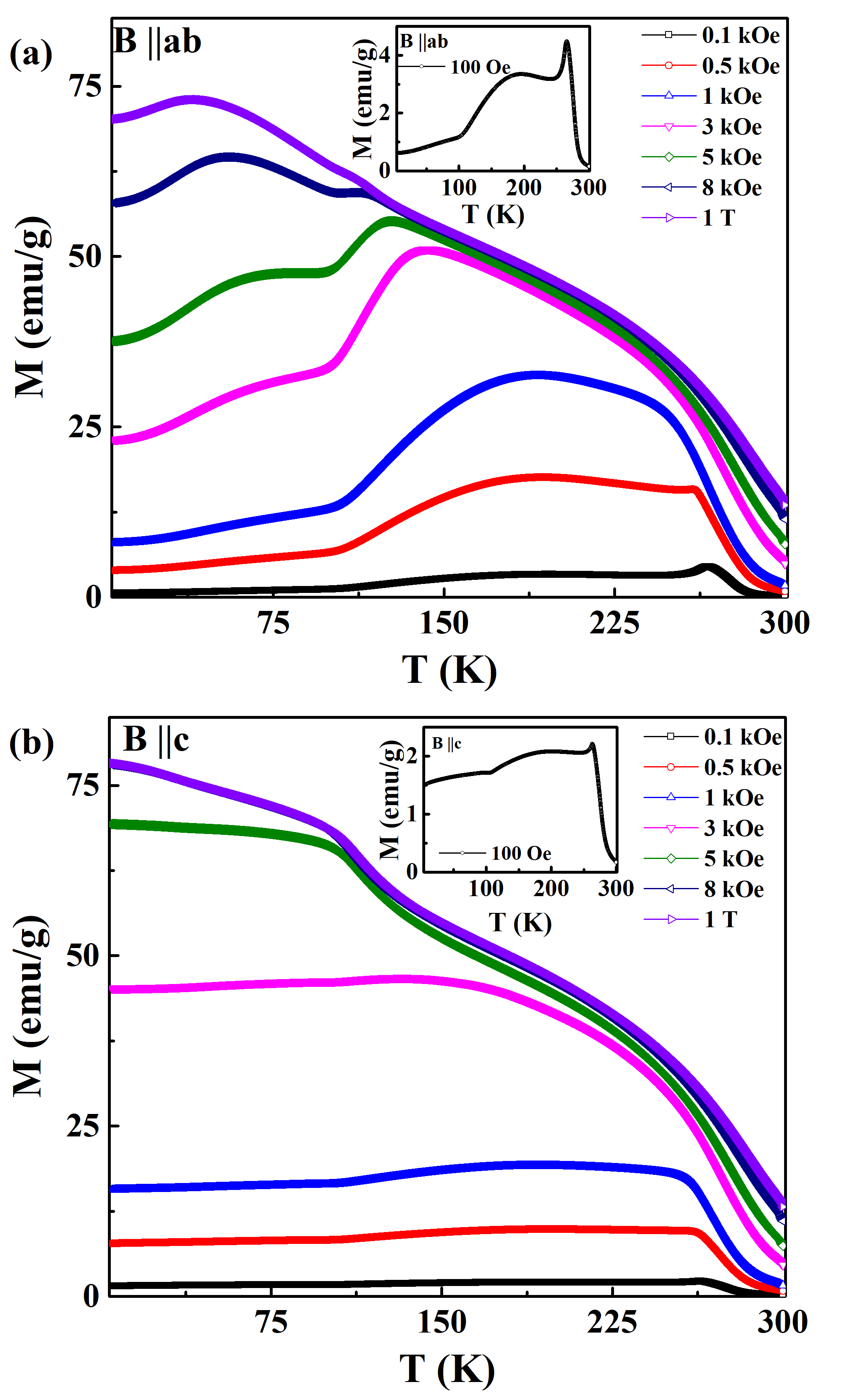}
\caption{(a) Temperature dependence of the dc magnetization for Fe$_4$GeTe$_2$ single crystal measured in zero-field-cooled condition for different magnetic fields applied parallel to the $ab$-plane (a) and  parallel to the $c$-axis (b). The insets show $M$($T$) curve at 100 Oe with $B\|ab$ and $B\|c$ configurations.}
\end{figure}
The temperature dependence of  magnetization for the Fe$_4$GeTe$_2$ bulk single crystal is shown in Figs. 1(a) and (b) for different applied fields parallel to the $ab$-plane ($B\|ab$) and $c$-axis ($B\|c$), respectively. Upon cooling, $M$ increases sharply at the PM to FM phase transition temperature, $T_C$=270 K, determined from the position of sharp minimum of the temperature derivative of the magnetization curve. The detailed analysis of magnetization and susceptibility data at and close to $T_C$ reveals that Fe$_4$GeTe$_2$ undergoes a second order phase transition at 270 K \cite{Mondal}. Remarkably, below $T_C$, the nature of  $M$($T$) curve for both $B\|ab$ and $B\|c$ configurations is very different from that one expects for a homogeneous FM system. At low field (100 Oe), the $M$($T$) curve for $B\|ab$ exhibits a sharp peak slightly below $T_C$ and $M$ does not increase monotonically with further decrease in $T$ below this peak but passes through a broad maximum around 195 K [Inset of Fig.1(a)]. The peak below $T_C$ is just visible above the background at 0.5 kOe field [Fig.1(a)]. On the other hand, the broad maximum shifts continuously toward lower temperature side with increase in field strength and  remains visible up to 1 T.  At $\sim$3 kOe, another maximum starts to appear around 65 K which also moves slowly toward lower temperature with increase in field and disappears above 1 T. For $B\|c$,  the $M$($T$) curve also exhibits a peak slightly below $T_C$ and the broad maximum around 195 K at 100 Oe [Inset of Fig.1(b)]. However, these features are  not as pronounced as  in the case of $B\|ab$ configuration and they either disappear or become very weak at a field of few hundred Oe [Fig.1(b)]. $M$ along this direction also decreases with decrease in $T$  but at a slower rate and for applied field only up to 3 kOe. Above 3 kOe, $M$ shows a sharp increase below 115 K.  We would like to mention that  at low fields, the $M$($T$) curve along both directions exhibits weak anomaly at 115 K. Though several anomalies displayed in $M$($T$) curve of Fe$_4$GeTe$_2$ are qualitatively similar to that reported for Fe$_{5-x}$GeTe$_2$, there are some important differences too\cite{Zhang}. The most important one is the value of $M$ at fixed $T$ and $B$. From the insets of Figs.1(a) and (b), it is clear that $M$ is larger at high temperatures for $B\|ab$ while this behavior reverses at low temperatures below $T_{SR}\sim$115 K. This suggests that a change in easy axis of magnetization takes place around $\sim$115 K. In contrast, the out-of-plane magnetization ($B\|c$) is smaller at all temperatures in Fe$_{5-x}$GeTe$_2$ \cite{Zhang}.

In order to track the temperature evolution of the Fe spin reorientation phenomenon, we have measured the field dependence of $M$ at different temperatures for $B\|ab$ and $B\|c$  as shown in Fig.2. Though,  at a given temperature, the $M$($B$) curves for both directions almost merge each other at high fields, the dependence of $M$ on $B$ in the low-field region is  extremely sensitive to direction of applied magnetic field.  At high temperature, $M$ is slightly larger for field $B\|ab$. However,  the difference  decreases with decrease in $T$ and disappears at around 115 K. At low temperature below 115 K, $M$ is larger for applied field parallel to $c$-axis. This suggests that the magnetic easy axis lies within the $ab$-plane at high temperature  and it rotates continuously toward the $c$-axis as $T$ approaches to $T_{SR}$  and the $c$-axis becomes the easy axis below $T_{SR}$. The spin reorientation transition has also been confirmed in few layers of Fe$_4$GeTe$_2$ via magnetic circular dichroism (MCD) measurement \cite{JSeo1}.
\begin{figure}
\includegraphics[width=0.5\textwidth]{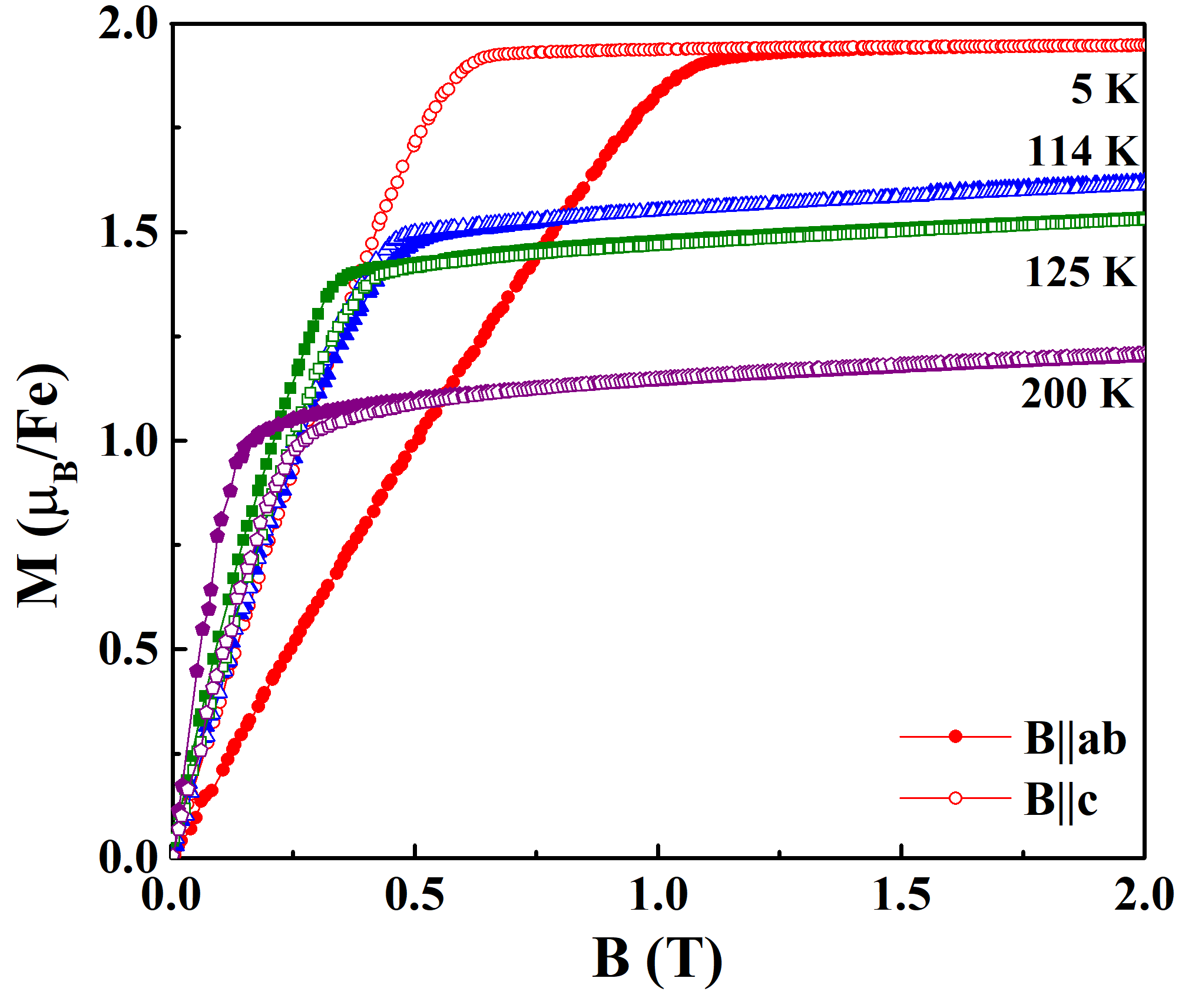}
\caption{Isothermal magnetization curves for the Fe$_{4}$GeTe$_2$ single crystal as a function of magnetic field at different temperatures measured in the vicinity of $T_{SR}$ and beyond for magnetic field  parallel to the  $c$-axis (open symbols) and parallel to the $ab$-plane (closed symbols).}
\end{figure}

For in-depth understanding the spin reorientation phenomenon in Fe$_4$GeTe$_2$, we have measured and analyzed the dependence of $M$ on $B$ in the vicinity of $T_{SR}$. Figures 3(a) and (b) show  isothermal $M$($B$) curves at few selected temperatures as representatives for $B\|c$ and $B\|ab$ configurations, respectively. The temperature and field evolution of $M$  are very different from that for a simple ferromagnet. Initially, $M$ increases sharply with increase in $B$ and then starts to saturate above a critical value of applied field ($B_s$). The value of $B_s$ depends on the crystallographic directions. It is clear from Fig.3(a) that $B_s$ is small along the $c$-axis and $M$ increases almost linearly with same slope below $B_s$ in a wide range of temperature. This linear behavior has been observed down to 2 K (not shown in Fig. 3).

\begin{figure}
\includegraphics[width=0.5\textwidth]{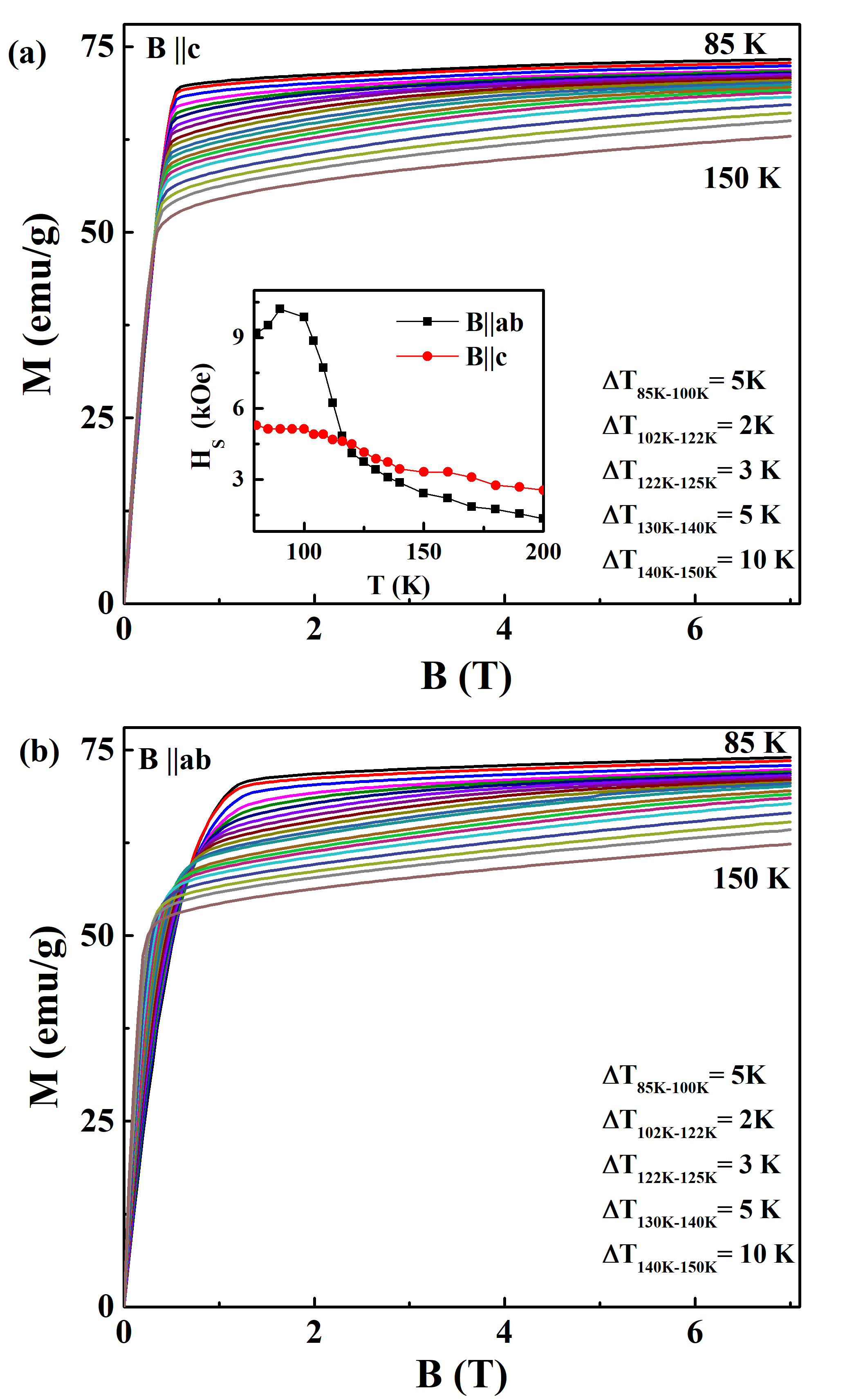}
\caption{Isothermal magnetization curves for the Fe$_{4}$GeTe$_2$ single crystal as a function of magnetic field at different temperatures measured in the vicinity of $T_{SR}$ for magnetic field parallel to the $c$-axis (a) and parallel to the $ab$-plane (b).}
\end{figure}

The nature of $M$($B$) curves for field parallel to $ab$-plane is very different. At low temperature, $M$ increases at a much slower rate with $B$ as compared to that for field along $c$-axis and exhibits a weak downward curvature at low fields [Fig.3(b)]. Though, $M$ increases at a slower rate at low temperatures, the rate  progressively increases with increase in $T$. Due to the increase of slope with temperature, the $M$($B$) curves along this direction cross each other. Also, the slope of $M$($B$) curve becomes larger than that for $c$-axis above $T_{SR}$. In Fe$_3$GeTe$_2$ and several vdW ferromagnets, $M$($B$) curves  exhibit such type of crossing below $T_C$, however, this phenomenon is weaker as compared to Fe$_4$GeTe$_2$ and the easy axis of magnetization does not change its direction with temperature \cite{Chen,YWang,Mondal1,Liu,Yan}. We have determined the temperature dependence of $B_s$ for $B\|c$ and $B\|ab$ configurations as shown in the inset of Fig.3(b). The value of saturation field $B_s$ has been estimated  from the intercept of two linear fits, one being a fit to the saturated regime at high fields and another being a fit to the low-field magnetization. The inset of Fig.3(b) shows that $B_s$ decreases very slowly with increase in temperature for $B\|c$, whereas  $B_s$ shows nonmonotonic $T$ dependence for $B\|ab$. For $B\|ab$, initially $B_s$  increases up to $T\sim$ 95 K and then decreases rapidly with further increase in $T$.  The $B_s$($T$) curves along two crystallographic directions cross each other at around $T_{SR}$ and $B_s$  becomes slightly smaller  along $ab$-plane above $T_{SR}$. The net anisotropy decreases with increase of temperature. This anisotropy in magnetization arises from magnetocrystalline anisotropy and shape anisotropy.
\begin{figure}
\includegraphics[width=0.5\textwidth]{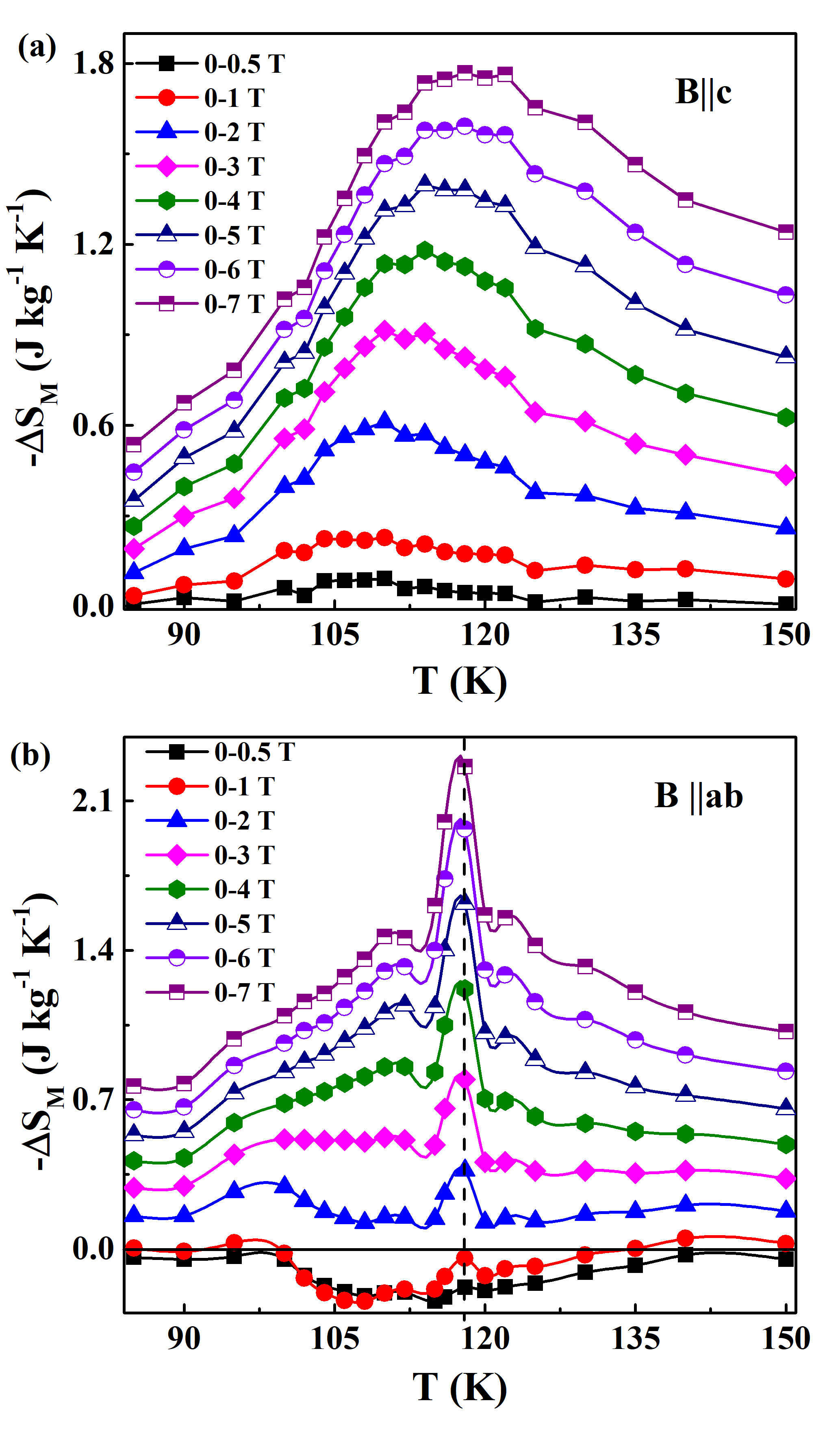}
\caption{Temperature dependence of magnetic entropy change -$\Delta S_M $ for Fe$_{4}$GeTe$_2$ around $T_{SR}$ determined from magnetization data (Fig.3) at different magnetic fields parallel to the $c$-axis (a) and parallel to the $ab$-plane (b). The dotted line is the spin reorientation transition temperature $T_{SR}$.}
\end{figure}

To explore the nature of magnetic interaction and magnetism in Fe$_4$GeTe$_2$ system, the magnetic entropy change has been calculated across the spin reorientation and FM transitions. $\Delta S_M $ of a system is related to the magnetization through the well-known thermodynamic Maxwell relation, $\Delta S_M=\int_{B_i}^{B_f}(\partial M/\partial T)_BdB$= $\sum_i \frac{M_{i+1}-M_{i}}{T_{i+1}-T_{i}}\Delta B_i$, where $M_i$ and $M_{i+1}$ are the initial magnetization values at temperatures $T_i$ and $T_{i+1}$, respectively for a field change of $\Delta B_i$.
As $\Delta S_M$ depends on ${\partial M}$/${\partial T}$ instead of $M$, a small change in  magnetization will be reflected clearly in $\Delta S_M$. Apart from the value, the sign of $\Delta S_M $ is also very important.  The macroscopic spin configuration of the system determines the sign of $\Delta S_M $. For these reasons, magnetocaloric effect has been extensively  used  to investigate the weak and unusual magnetic phase transitions and the coexistence of different magnetic phases in the studied compound \cite{Jamal,Kohama,Lampen,Aoki,Yan1,Nandi}.
\begin{figure}
\includegraphics[width=0.5\textwidth]{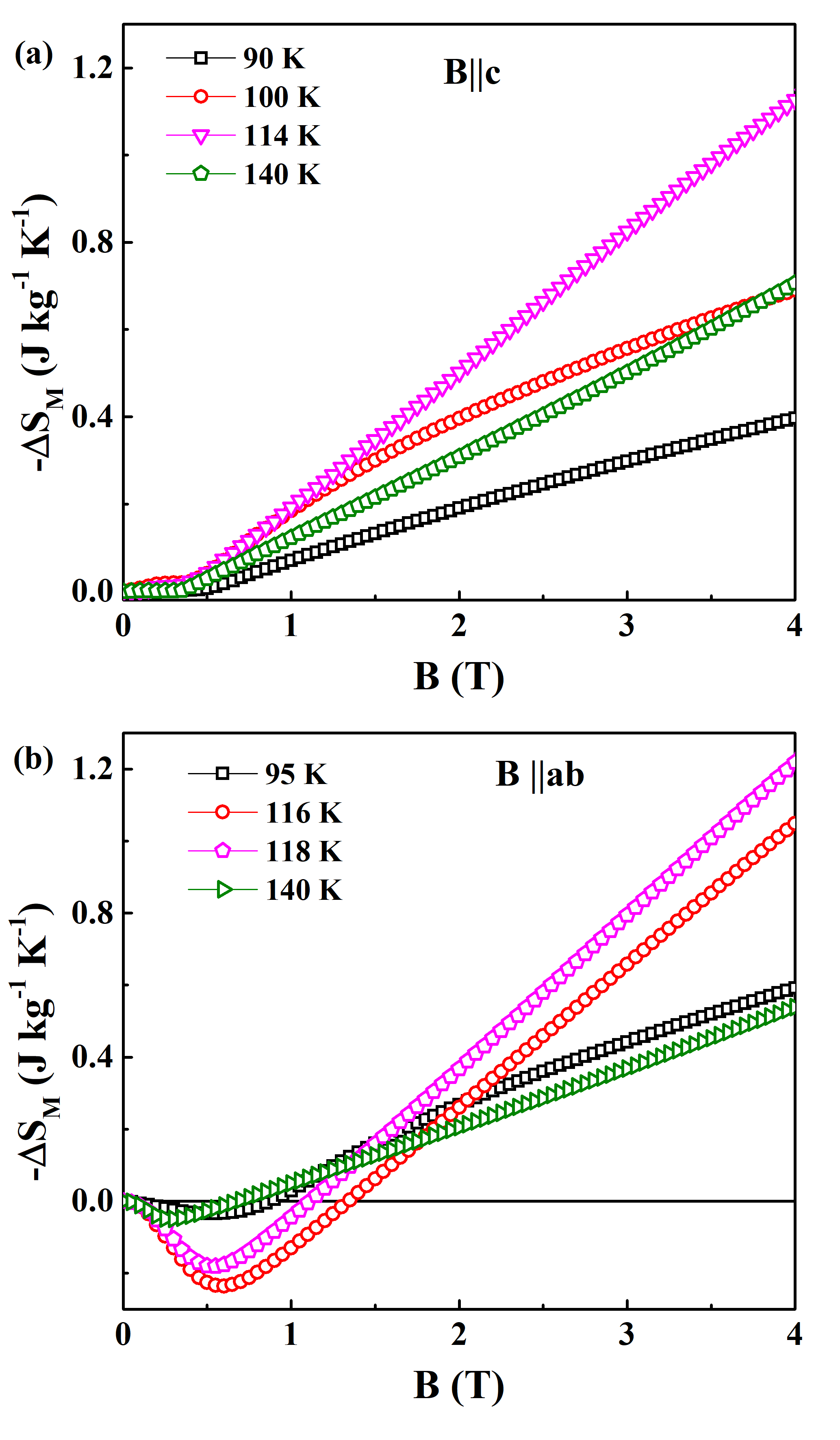}
\caption{Field dependence of magnetic entropy change -$\Delta S_M $ for Fe$_{4}$GeTe$_2$ around $T_{SR}$ estimated from the magnetization data (Fig.3) at different temperatures parallel to the $c$-axis (a) and parallel to the $ab$-plane (b).}
\end{figure}

$\Delta S_M $ determined at different fields up to 7 T is shown in Figs. 4(a) and (b) as a function of temperature for  $B\|c$ and  $B\|ab$ directions, respectively. -$\Delta S_M $($T$) along $c$-axis exhibits a broad maximum around $T_{SR}$ which is very similar to that observed in several magnetic materials in the vicinity of their PM to FM/AFM phase transition. The observed nature of $\Delta S_M $($T$) curve clearly suggests that the spin reorientation is a thermodynamic phase transition at $T_{SR}$. Apart from this, $\Delta S_M $($T$) also displays several anomalies those are important for understanding the nature of magnetic interaction in the present system. For $B\|c$, $\Delta S_M $ is negative at all fields. $\Delta S_M $ for $B\|ab$ is also negative for $B\geq$ 2 T but the nature of -$\Delta S_M $($T$) curve is very unusual along this direction and it progressively changes with the increase in field strength. -$\Delta S_M$($T$) exhibits a peak at $T_{SR}$ which becomes very sharp at high field. At low fields $B<$ 2 T, $\Delta S_M$ is positive  in the temperature range 90 to 140 K. We have also determined the dependence of $\Delta S_M$ on $B$ close to $T_{SR}$. Figures 5(a) and (b) show -$\Delta S_M$($B$) curves at different temperatures for both directions of applied field. $\Delta S_M$ is negative and it decreases continuously with field for $B\|c$. $\Delta S_M$ decreases very slowly at low fields below $B_s$ but decreases at a much faster rate above $B_s$. On the other hand, for $B\|ab$, $\Delta S_M$($B$) shows a non-monotonic dependence on $B$  and is positive below a critical value of applied field which depends on $T$.  Close to $T_{SR}$, $\Delta S_M$ is maximum ($\sim$0.25 Jkg$^{-1}$K$^{-1}$) at 0.7 T  and changes its sign  at around 1.4 T.

\begin{figure}
\includegraphics[width=0.5\textwidth]{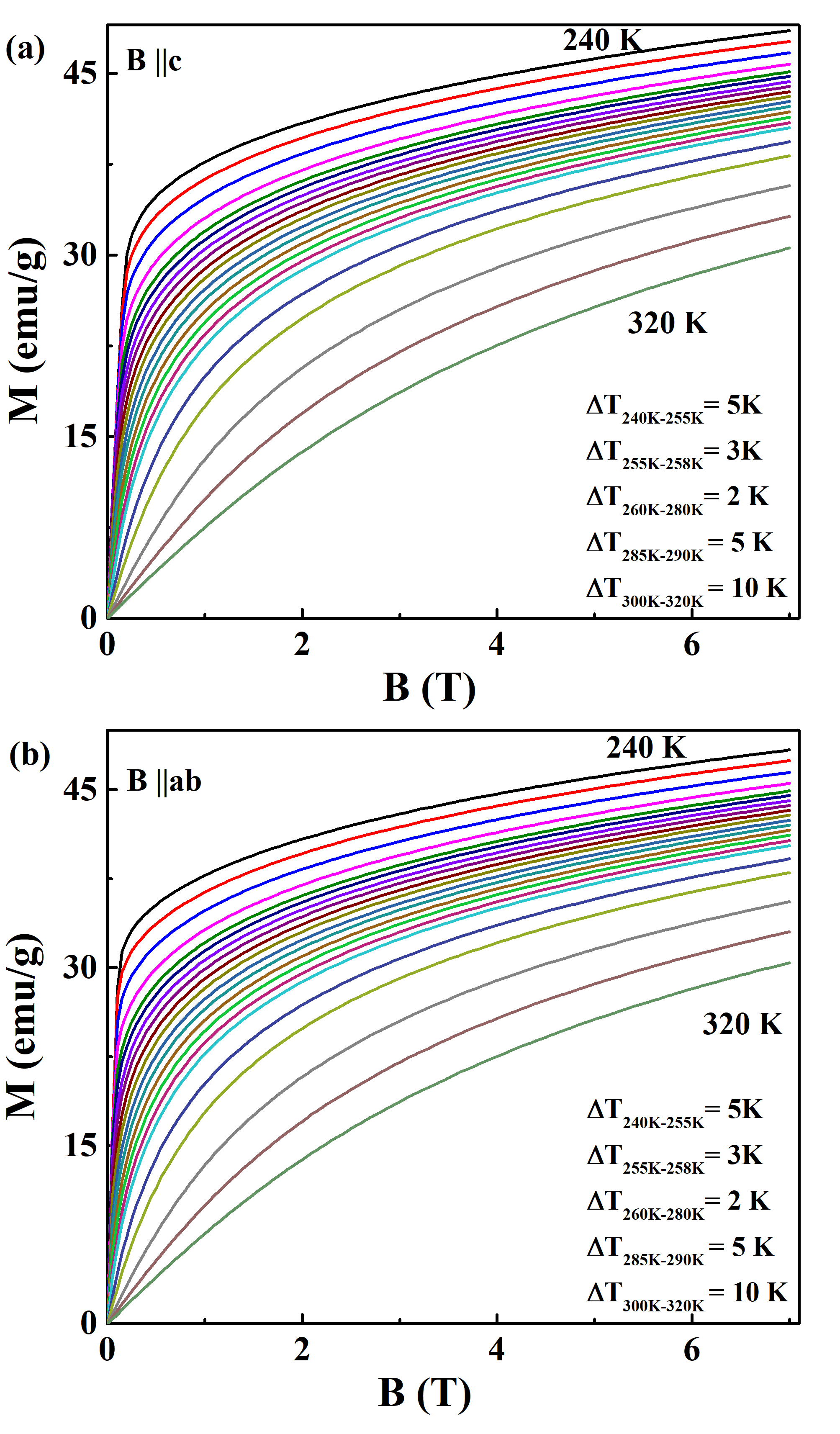}
\caption{Isothermal magnetization for Fe$_{4}$GeTe$_2$ as a function of magnetic field at different temperatures measured in the vicinity of $T_{C}$ for magnetic field parallel to the $c$-axis  (a) and parallel to the $ab$-plane (b).}
\end{figure}
\begin{figure}
\includegraphics[width=0.5\textwidth]{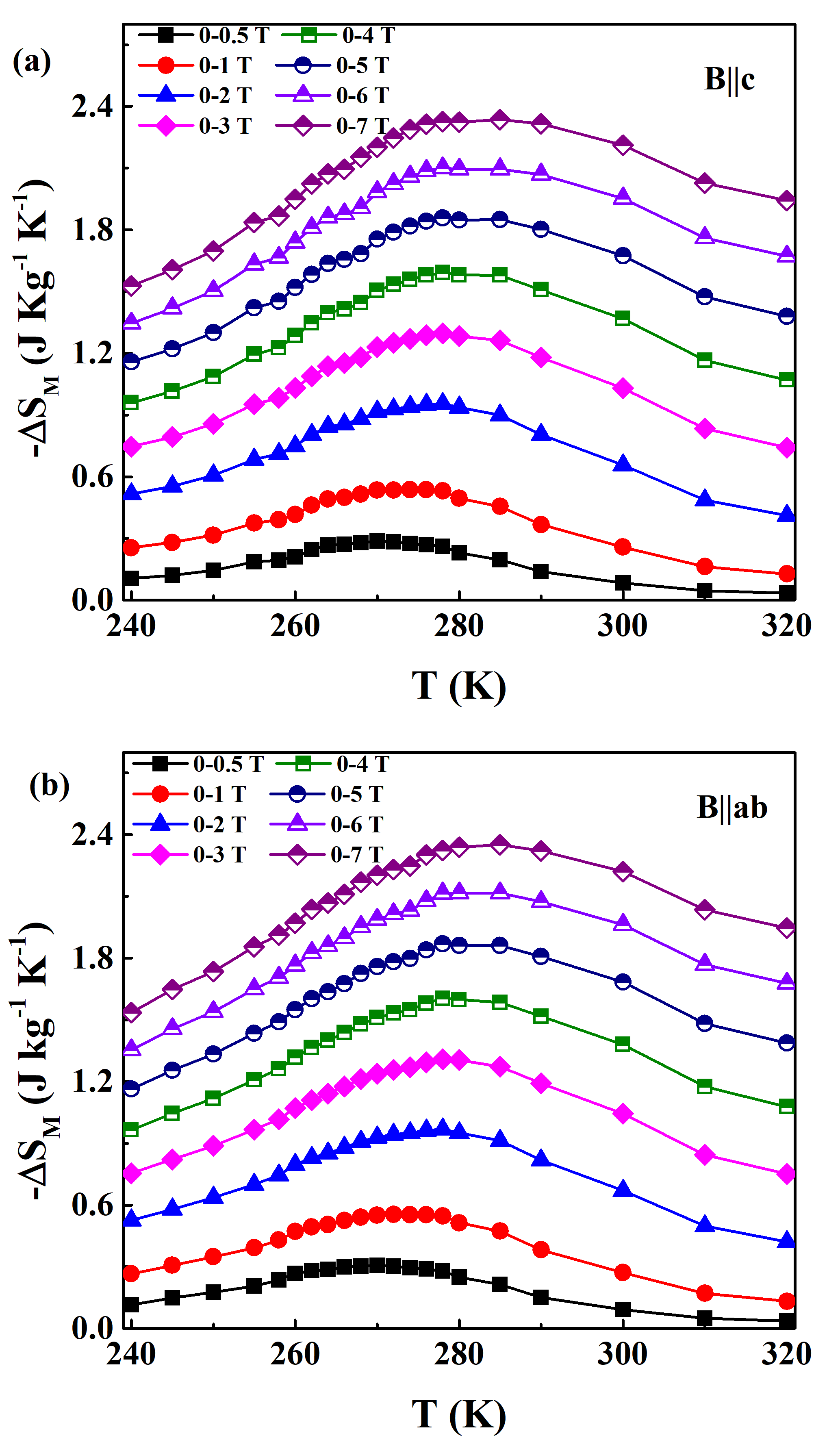}
\caption{Temperature dependence of magnetic entropy change -$\Delta S_M $ for Fe$_{4}$GeTe$_2$ around $T_{C}$ determined from the magnetization data (Fig.6) at different magnetic fields parallel to the $c$-axis (a) and parallel to the $ab$-plane (b).}
\end{figure}

We have recorded isothermal magnetization across the FM-PM transition in Fe$_4$GeTe$_2$ to calculate $\Delta S_M $. Figures 6(a) and (b) show the field dependence of magnetization in the temperature range 240-320 K for $B\|c$ and $B\|ab$, respectively. The nature of $M$($B$) curves in both directions is very similar to that one expects for a FM system. At a given field, $M$ decreases monotonically with increasing temperature. The temperature dependence of $\Delta S_M $ is shown in Figs.7(a) and (b) for $B\|c$ and $B\|ab$, respectively. $\Delta S_M $ is found to be negative over the measured temperature range and the nature of $\Delta S_M $($T$) curve for both directions is typical of a FM system. As expected, the $\Delta S_M $($T$) curve shows a broad peak around $T_C$ and the height of peak increases monotonically with field.

Positive value of $\Delta S_M $ with field along $ab$-plane implies that applied magnetic field creates more disorder in the system. This behavior is very unusual  because external magnetic field suppresses disorder in a FM system. Nevertheless, several FM vdW compounds have been reported to exhibit small positive $\Delta S_M $ below $T_C$ when the field is applied along the hard axis of magnetization \cite{Liu,Yan}. Several antiferromagnetic systems, in which $M$($B$) curves cross each other at low fields and low temperature, exhibit positive $\Delta S_M $ \cite{Midya,Midya1,Nandi}. When magnetic field is applied, the magnetic moment fluctuation is enhanced in one of the two AFM sublattices which is antiparallel to $B$. As $B$ increases, more and more spins in the antiparallel sublattice orient along the field direction. This, in turn, enhances the spin disordering and, hence $\Delta S_M $  becomes positive. Normally, this trend will continue up to a certain field. As the majority of spins in the antiparallel sublattice orient along the field direction at a moderate field, the system is expected to show negative $\Delta S_M $. Thus one may ask whether there is weak AFM interaction along the hard axis of magnetization due to  the existence of complex magnetic phases in the low-temperature region. However, the exact mechanism behind this intriguing magnetic properties is not yet known. In this context, we would like to mention that detailed magnetotransport measurements on Fe$_3$GeTe$_2$ single crystal suggests non-collinear nature of spin configuration along hard axis below $B_s$ \cite{YWang}. Scanning tunneling microscopy study in Fe$_{5-x}$GeTe$_2$, which is also showing several anomalies in $M$($T$) curve, reveals ordering of Fe(1) layer and inhomogeneous nature of magnetism due to the formation of stripe phase \cite{Ly}. It has been further argued that the ordering of Fe(1) breaks the inversion symmetry of crystal  and as a result, strong antisymmetric  Dzyalosinskii-Moriya (DM) interaction develops in the system and a helimagnet ground state  arises  due to the competition between the DM interaction and ferromagnetic exchange. The sharp peak below $T_C$ is attributed to the helimagnet. Further microscopic study and neutron scattering  experiments will be useful for understanding the nature of magnetic ground state of Fe$_4$GeTe$_2$.

\section{Conclusion}
In conclusion, we have studied the magnetic properties of the quasi two-dimensional itinerant ferromagnet Fe$_4$GeTe$_2$ across the FM-PM transition and spin reorientation transition with applied field parallel to $ab$-plane and $c$-axis. Unlike a typical ferromagnet, both temperature and magnetic field dependence of magnetization is very complex in nature possibly due to the presence of several competitive magnetic interactions. Below $T_C$, $M$($T$) curves display multiple maxima and their nature and position are very sensitive to the direction and strength of the applied field. Above $T_{SR}$=115 K, the easy axis of magnetization lies within the $ab$-plane and it rotates along the $c$-axis below $T_{SR}$. Well below $T_C$ and saturation field, $M$ along the $c$-axis increases sharply with a linear slope which is independent of $T$ while $M$($B$) curves  along $ab$-plane show sublinear behavior and they cross each other due to the increase of slope with increase in $T$. Interestingly, $\Delta S_M $ is highly anisotropic close to $T_{SR}$. With field along $c$-axis, $\Delta S_M $ is negative and -$\Delta S_M $($T$) shows a broad peak around $T_{SR}$ similar to the case of FM-PM transition. In contrast, with field parallel to $ab$-plane, $\Delta S_M $ is positive below 1.4 T  in the range 90-140 K and -$\Delta S_M $($T$) shows a much sharper peak at $T_{SR}$. The nature of both $M$($B$) and $\Delta S_M $($T$) in the vicinity of $T_{C}$ is similar to a typical FM. The observation of peak around $T_{SR}$ in -$\Delta S_M $($T$) suggests that the spin reorientation transition is a thermodynamic phase transition.\\

\section{ACKNOWLEDGMENTS}
The authors are very much thankful to Dr. Dipten Bhattacharya for his valuable assistance during the characterization of the sample. The authors also would like to thank A. Paul for his help during the measurements.
\newpage
\bibliographystyle{apsrev4-1}

\begin{thebibliography}{99}

\bibitem {AKGeim} A. K. Geim and I. V. Grigorieva, Nature (London) \textbf{499}, 419 (2013).

\bibitem {TSong} T. Song, X. Cai, M. W. Tu, X. Zhang, B. Huang, N. P. Wilson, K. L. Seyler, L. Zhu, T. Taniguchi, K. Watanabe, M. A. McGuire, D. H. Cobden, D. Xiao, W. Yao, and X. Xu, Science \textbf{360}, 1214 (2018).

\bibitem {DKlein} D. R. Klein, D. MacNeill, J. L. Lado, D. Soriano, E. Navarro-Moratalla, K. Watanabe, T. Taniguchi, S. Manni, P. Canfield, J. Fern\'{a}ndez-Rossier, and P. Jarillo-Herrero, Science \textbf{360}, 1218 (2018).

\bibitem {CGong} C. Gong, L. Li, Z. Li, H. Ji, A. Stern, Y. Xia, T. Cao, W. Bao, C. Wang, Y. Wang, Z. Q. Qiu, R. J. Cava, S. G. Louie, J. Xia, and X. Zhang, Nature (London) \textbf{546}, 265 (2017).

\bibitem{Mcg} M. A. McGuire, H. Dixit, V. R. Cooper, and B. C. Sales, Chem. Mater. \textbf{27}, 612 (2015).

\bibitem {BHuang} B. Huang, G. Clark, E. Navarro-Moratalla, D. R. Klein, R. Cheng, K. L. Seyler, D. Zhong, E. Schmidgall, M. A. McGuire, D. H. Cobden, W. Yao, D. Xiao, P. Jarillo-Herrero, and X. Xu, Nature (London) \textbf{546}, 270 (2017).

\bibitem {Burch} K. S. Burch, D. Mandrus, and J.-G. Park, Nature \textbf{563}, 47 (2018).

\bibitem {Casto} L. D. Casto, A. J. Clune, M. O. Yokosuk, J. L. Musfeldt, T. J. Williams, H. L. Zhuang, M.-W. Lin,
K. Xiao, R. G. Hennig, B. C. Sales, J.-Q. Yan, and D. Mandrus, APL Mater. \textbf{3}, 041515 (2015).

\bibitem {Carte} V. Carteaux, D. Brunet, G. Ouvrard, and G. Andre, J. Phys. Condens. Matter \textbf{7}, 69 (1995).

\bibitem {Tsubo} I. Tsubokawa,  J. Phys. Soc. Jpn. \textbf{15}, 1664 (1960).

\bibitem {Deise} H.-J. Deiseroth, K. Aleksandrov, C. Reiner, L. Kienle, and R. K. Kremer, Eur. J. Inorg. Chem. 2006, 1561 (2006).

\bibitem {Siber} B. Siberchicot, S. Jobic, V. Carteaux, P. Gressier, and G. Ouvrard, J. Phys. Chem. \textbf{100}, 5863 (1996).

\bibitem {Deng} Y. Deng, Y. Yu, Y. Song, J. Zhang, N. Z. Wang, Z. Sun, Y. Yi, Y. Z. Wu, S. Wu, J. Zhu, J. Wang, X. H. Chen, and Y. Zhang, Nature \textbf{563}, 94 (2018).

\bibitem {Chen} B. Chen, J. Yang, H. Wang, M. Imai, H. Ohta, C. Michioka, K. Yoshimura, and  M. Fang, J. Phys. Soc. Jpn. \textbf{82}, 124711 (2013).

\bibitem {KKim} K. Kim, J. Seo, E. Lee, K. T. Ko, B. S. Kim, B. G. Jang, J. M. Ok, J. Lee, Y. J. Jo, W. Kang, J. H. Shim, C. Kim, H. W. Yeom, B. Il Min, B. J. Yang, and J. S. Kim, Nat. Mater. \textbf{17}, 794 (2018).

\bibitem {YWang} Y. Wang, C. Xian, J. Wang, B. Liu, L. Ling, L. Zhang, L. Cao, Z. Qu, and Y. Xiong, Phys. Rev. B \textbf{96}, 134428 (2017).

\bibitem {CTan} C. Tan, J. Lee, S.-G. Jung, T. Park, S. Albarakati, M. R. Partridge, J.and Field, D. G. McCulloch, L. Wang, and C. Lee, Nat. Commun. \textbf{9}, 1554 (2018).


\bibitem {AFMay1} A. F. May, D. Ovchinnikov, Q. Zheng, R. Hermann, S. Calder, B. Huang, Z. Fei, Y. Liu, X. Xu, and M. A. McGuire, ACS Nano \textbf{13}, 4436 (2019).

\bibitem {AFMay2} A. F. May, C. A. Bridges, and M. A. McGuire, Phys. Rev. Mater. \textbf{3}, 104401 (2019).

\bibitem {JSeo1} J. Seo, D. Y. Kim, E. S. An, K. Kim, G.-Y. Kim, S.-Y. Hwang, D. W. Kim, B. G. Jang, H. Kim, G. Eom, S. Y. Seo, R. Stania, M. Muntwiler, J. Lee, K. Watanabe, T. Taniguchi, Y. J. Jo, J. Lee, B. I. Min, M. H. Jo, H. W. Yeom, S.-Y. Choi, J. H. Shim, and J. S. Kim, Sci. Adv. \textbf{6}, eaay8912 (2020).

\bibitem{Fei} Z. Fei, B. Huang, P. Malinowski, W.Wang, T. Song, J. Sanchez, W. Yao, D. Xiao, X. Zhu, A. F. May, W. Wu, D. H. Cobden, J. H. Chu, and X. Xu, Nat. Mater. \textbf{17}, 778 (2018).

\bibitem{Ding} B. Ding, Z. Li, G. Xu, H. Li, Z. Hou, E. Liu, X. Xi, F. Xu, Y. Yao, and W. Wang, Nano Lett. \textbf{20}, 868 (2020).

\bibitem{Zhang} H. Zhang, R. Chen, K. Zhai, X. Chen, L. Caretta, X. Huang, R. V. Chopdekar, J. Cao, J. Sun, J. Yao, R. Birgeneau, and R. Ramesh, Phys. Rev. B \textbf{102}, 064417 (2020).

\bibitem{Stahl} J. Stahl, E. Shlaen, and D. Johrendt, Z. Anorg. Allg. Chem. \textbf{644}, 1923 (2018).

\bibitem{Gao} Y. Gao, Q. Yin, Q. Wang, Z. Li, J. Cai, T. Zhao, H. Lei, S. Wang, Y. Zhang, and B. Shen, Adv. Mater. \textbf{32}, 2005228 (2020).

\bibitem{Mondal} S. Mondal, N. Khan, S. M. Mishra, B. Satpati, and P. Mandal, Phys. Rev. B \textbf{104}, 094405 (2021).

\bibitem {JSeo2} J. Seo, E. S. An, T. Park, S.-Y. Hwang, G.-Y. Kim, K. Song, W.-S. Noh, J. Y. Kim, G. S. Choi, M. Choi, E. Oh, K. Watanabe, T. Taniguchi, J. H. Park, Y. J. Jo, H. W. Yeom, S.-Y. Choi, J. H. Shim, and J. S. Kim, Nat. Commun. \textbf{12}, 2844 (2021).

\bibitem{Mondal1} S. Mondal, M. Kannan, M. Das, L. Govindaraj, R. Singha, B. Satpati, S. Arumugam, and P. Mandal, Phys. Rev. B \textbf{99}, 180407(R) (2019).

\bibitem {Liu} Y. Liu and C. Petrovic, Phys. Rev. B \textbf{97}, 174418 (2018).

\bibitem{Yan} J. Yan, X. Luo, F. C. Chen, J. J. Gao, Z. Z. Jiang, G. C. Zhao, Y. Sun, H. Y. Lv, S. J. Tian, Q. W. Yin, H. C. Lei, W. J. Lu, P. Tong, W. H. Song, X. B. Zhu, and Y. P. Sun, Phys. Rev. B \textbf{100}, 094402 (2019).

\bibitem{Jamal} Sk Jamaluddin, S. K. Manna, B. Giri, P. V. Prakash Madduri, S. S. P. Parkin, and A. K. Nayak, Adv. Funct. Mater. \textbf{29}, 1901776 (2019).

\bibitem{Kohama} Y. Kohama, H. Ishikawaa , A. Matsuoa , K. Kindoa , N. Shannon, and Z. Hiroi, Proc. Natl. Acad. Sci. USA \textbf{116}, 10686 (2019).

\bibitem{Lampen} P. Lampen, N. S. Bingham, M. H. Phan, H. Srikanth, H. T. Yi, and S. W. Cheong, Phys. Rev. B \textbf{89}, 144414 (2014).

\bibitem{Aoki} H. Aoki, T. Sakakibara, K. Matsuhira, and Z. Hiroi, J. Phys. Soc. Jpn. \textbf{73}, 2851 (2004).

\bibitem{Yan1}  L. Q. Yan, S. H. Chun, Y. Sun, K. W. Shin, B. G. Jeon, S. P. Shen, and K. H. Kim, J. Phys.: Condens. Matter\textbf{25}, 256006 (2013).

\bibitem{Nandi} M. Nandi and P. Mandal, J. Appl. Phys. \textbf{119}, 133904 (2016).

\bibitem{Midya} A. Midya, S. N. Das, P. Mandal, S. Pandya, and V. Ganesan, Phys. Rev. B \textbf{84}, 235127 (2011).

\bibitem{Midya1} A. Midya, N. Khan, D. Bhoi, and  P. Mandal, Appl. Phys. Lett. \textbf{103}, 092402 (2013).

\bibitem{Ly} T. T. Ly, J. Park, K. Kim, H.-B. Ahn, N. J. Lee, K. Kim, T.-E. Park, G. Duvjir, N. H. Lam, K. Jang, C.-Y. You, Y. Jo, S. K. Kim, C. Lee, S. Kim, J. Kim, Adv. Funct. Mater. \textbf{31}, 2009758 (2021).
\end{thebibliography}

\end{document}